\documentclass[11pt]{article}

\usepackage{fullpage,amsfonts,amssymb,amsthm,amsmath}
\usepackage{graphicx}
\graphicspath{ {./images/} }
\usepackage[colorlinks, bookmarksnumbered, linkcolor=blue, citecolor=blue, urlcolor=green]{hyperref}
\usepackage{geometry}
\usepackage{algorithm}

\usepackage[noend]{algpseudocode}

\usepackage{amsmath, amsthm, amssymb}
\usepackage{url}
\usepackage{physics}
\usepackage{graphicx,lipsum}
\graphicspath{ {./downloads/} }

\usepackage{pythonhighlight}

\makeatletter

\newtheorem{theorem}{Theorem}
\newtheorem{corollary}{Corollary}
\newtheorem{fact}{Fact}

\theoremstyle{definition}

\makeatletter
  \renewcommand{\ALG@name}{Protocol}
  \makeatother
\usepackage{float}
\usepackage{caption}
\usepackage{subcaption}
\usepackage{authblk}
\usepackage{mathtools}

\usepackage{tabularx}

\newcounter{protocol}

\title{Automated Market Makers for Cross-chain DeFi and Sharded Blockchains}

\author[1]{Jon Michael Aanes\thanks{jon.michael.aanes@partisia.com}}
\author[1,2]{Jesper Balman Gravgaard\thanks{jg@partisia.com}}
\author[1]{Peter Bro Miltersen\thanks{peter.bro.miltersen@partisia.com}}
\author[1,2,3]{Kurt Nielsen\thanks{kn@partisia.com}}
\author[4, 5]{Mohsen Pourpouneh\thanks{m.pourpouneh@maastrichtuniversity.nl}}

\affil[1]{Partisia}
\affil[2]{Partisia Blockchain Foundation}
\affil[3]{University of Copenhagen, Copenhagen, Denmark}
\affil[4]{University of Oxford, Oxford, United Kingdom}
\affil[5]{Maastricht University, Maastricht, The Netherlands}
\begin{document}
\maketitle
	\maketitle
	\noindent
	\rule[0.5ex]{1\columnwidth}{1pt}
	\begin{abstract}
          We consider Uniswap-like automated market makers, and, specifically, constant product liquidity pools, operating on blockchains.
          An important feature of Uniswap is the ability for a trader to carry out a sequence of asset swaps atomically, without other traders changing the prices along the way. This atomic-execution feature is not immediately available in cross-chain or sharded blockchain settings, where different liquidity pools are distributed across different chains or shards. Our contribution is a description and suggested implementation of a new functionality that might be added to individual liquidity pools, the {\em lock-swap}. The lock-swap enables a trader to get a guarantee for the price associated with a swap but only decide later whether or not to carry out the swap. Applied across several liquidity pools, it guarantees the trader assured prices for all swaps in a swap sequence and lets these prices inform the trader's decision about whether or not to carry out the sequence, thus essentially giving the trader the same benefits an atomic execution of the sequence would have provided him. However, in contrast to an atomic execution, our functionality does not prevent other traders from doing swaps during the time where the sequence is planned and possibly carried out. Nor does it prevent liquidity providers from adding or removing liquidity to and from the liquidity pool in that time period.
	\end{abstract}
	
	\noindent
	\rule[0.5ex]{1\columnwidth}{1pt}

\section{Introduction}

Automated Market Makers (AMMs) are an important part of DeFi and a major driver of the Web $3.0$ adoption across different blockchain platforms. 
The initial motivation of AMM  was the opportunity to exchange crypto assets without direct interaction and matching of buyers or sellers.
An AMM solution is essentially one or more smart contracts safeguarded by the underlying blockchain platform. 
Ethereum has been the most used blockchain platform for AMM solutions, a pioneer being Uniswap
\cite{adamsuniswapv1, adamsuniswapv2, formal}, who introdued the now ubiquitous {\em constant product liquidity pool}.  Recent developments take this one step further and run AMMs across independent blockchain platforms, i.e., cross-chain DeFi.

In this paper, we suggest a method of emulating an important feature of Uniswap liquidity pools in a cross-chain setting, namely {\em the atomic execution of sequences of swaps}.  
To understand the importance of this feature, consider the following example: On Uniswap, an arbitrageur can swap asset $A$ to asset $B$, then swap asset $B$ to asset $C$ and finally swap asset $C$ to asset $A$, potentially completing a profitable arbitrage cycle without other traders interfering. Note, however, that if other traders were to interfere with their own swaps during the execution of such a planned cycle, then the arbitrageur might find after swapping $A$ to $B$ and $B$ to $C$ that the price he expected for the swap from $C$ to $A$ had changed and that his trade sequence would now lead to a loss rather than a profit if he were to carry out the last swap. Thus, his arbitrage might fail. In a cross-chain situation with liquidity pools distributed on different chains and no possibility of atomic execution of several swaps, this might indeed be the situation.

The cross-chain situation is similar to the most efficient sharding model that essentially operates like independent blockchains where transactions are automatically off-loaded across different shards in a way that favors unlimited parallelisation i.e. asynchronous and concurrent execution. Although very few blockchain networks comes with such real scalable sharding there is a few, such as the Partisia Blockchain\footnote{\url{https://partisiablockchain.gitlab.io/documentation/}}. Examples of other sharded blockchains are Elastico \cite{luu2016secure}, OmniLedger \cite{kokoris2018omniledger}, and RapidChain  \cite{zamani2018rapidchain}. In such settings, we encounter a similar situation to the above when different liquidity pools are on different shards: We pay for the scalability and the parallelization by not being able to carry out a sequence of swaps associated with pools in various shards atomically. 

This paper mitigates the situation described above by introducing a functionality one can add to individual Uniswap-style constant product liquidity pools. The functionality allows a trader to complete sequences or cycles of swaps at prices that are {\em guaranteed in advance}, just as they would have been if the swaps had been carried out atomically. We call our functionality a ``lock-swap''. A lock-swap gives a trader (for instance, an arbitrageur) a quote and an option for a particular swap, but does not require him to carry out the swap immediately. By collecting such options for all swaps in a desired sequence or cycle, and only then executing the swaps if the prices are collectively deemed desirable and canceling them if they are not, the arbitrageur has effectively emulated a Uniswap style atomic swap sequence. However, in contrast to Uniswap,  our solution does {\em not} prevent other traders from carrying out swaps (including lock-swaps). Nor does it prevent liquidity providers from providing and reclaiming liquidity while the arbitrageur works on his cycle (so the lock-swap locks the prices, but not the pool itself).

The paper is organized as follows. In section \ref{classical}, we review classical liquidity pools and their most important properties for our context. These properties turn out to be {\em asset positivity} and {\em product preservation}. In section \ref{npp}, we make a step towards the lock-swap enabled pool by introducing a functionality that we consider interesting in its own right; the ability to add liquidity to a constant product pool in arbitrary proportions. In section \ref{lockswap}, we define the lock-swap mechanism in an abstract way, by stating its required properties. In section \ref{retroactive}, we present our main construction acheiving these properties, the {\em retroactive liquidity pool}. In section \ref{python}, we describe a practical python prototype implementation of our construction. In section \ref{restricted}, we describe natural restricted settings that make our construction even more practical. We conclude in section \ref{discuss} by briefly and informally discussing the economic aspects of our construction.

\section{Classical liquidity pools}
\label{classical}

 Decentralized exchanges (DeX) are smart contracts that allows traders to swap assets without the need for any third party or broker. A well studied category of DeXs are the so called automated market makers (AMMs). To execute swaps AMMs use a bonding curve that determines the price of the asset (with respect to each other) purely based on the amount of asset (liquidity) locked in the smart contract. In its  simplest form, an AMM has a \textit{liquidity pool} for a pair of assets A (say, Bitcoin) and B (say, Ethereum). Traders can exchange asset A for asset B by depositing an amount of asset A in the smart contract and receiving an amount of asset B in return. Anyone can add liquidity by depositing A and B assets to the pool. To incentivize the liquidity providers for providing liquidity into the AMM, for each trade the pool charges a fee for every swap being executed, which is distributed (proportionally)  among those who provided liquidity. Liquidity providers are given liquidity tokens to keep track of their share of the pool. The AMMs use a deterministic pricing rule  to provide the exchange ratio (i.e., the price) between the two assets, hence they are also called ``Constant Function Market Makers'' \cite{angeris2020improved}. Uniswap \cite{adamsuniswapv1} is one of the most prominent implementation of such models that uses a  constant product function to determine the price between the pair of assets. Other examples includes  Stable Swap \cite{egorov2019stableswap}, and Balancer \cite{martinelli2019balancer}.

  
Formally, we are going to model such a classical constant product liquidity pool as a data structure $p$ with certain properties.

We first need an auxillary, very simple, concept. A {\em liquidity token portion} is given by a single positive real number: the {\em amount} of the portion. A liquidity token portion is intended to model a stash of liquidity tokens for a liquidity pool, with a portion typically minted as response to addition of liquidity to the pool. The amount of the portion simply represents the number of tokens in the portion. However, we consider two liquidity token portions different, even if their amounts are the same (as we need to keep track of particular portions in the statements and proofs). 

The state of a classical liquidity pool $p$ is at any point in time given by an {\em asset amount pair} $(a,b)$ of two positive real numbers $a > 0$ and $b > 0$ and another positive real number $z$. Here, $a$ represents an amount of an asset A (say, Bitcoin) and $b$ represents an amount of an asset B (say, Ethereum). A deployed liquidity pool will literally {\em hold} these assets (by being in possession of the private keys for blockchain accounts containing the assets), but the details of this aspect is largely ignored in our formal model. The number $z$ represents the total number of liquidity tokens already minted and not yet burned. The data structure is accessed using the following operations:
\begin{itemize}
\item{} $t \leftarrow p.\mbox{\bf init}(\Delta a, \Delta b)$, where $\Delta a, \Delta b$ are positive real numbers. This initializes the pool to $(a,b) = (\Delta a, \Delta b)$ and $z = 1$ and returns a liquidity token portion $t$ of amount 1 (the amount is in fact arbitrary, we use 1 for simplicity). This operation is only permitted once and must be the first operation performed.
\item{}$\Delta b \leftarrow p.\mbox{\bf swapAtoB}(\Delta a)$, where $\Delta a$ is a positive real number. A {\em negative} real number $\Delta b$ is given as output. If the current state of the pool is $(a,b)$, the following inequality must be satisfied:
  \begin{equation}
    \label{increase}
    (a + \Delta a)(b + \Delta b) \geq ab.
  \end{equation}
  As a result of the operation, the pool changes state from $(a,b)$ to $(a + \Delta a, b + \Delta b)$ with $z$ left unchanged. This models a swapper using the pool to exchange $\Delta a$ units of asset A to $-\Delta b$ units of asset B. The reason that (\ref{increase}) is an inequality rather than an equation is that we expect the pool to charge the swapper a fee. We put no constraint on the fee, except that it must be non-negative. If the fee is 0 so that (\ref{increase}) becomes an equation, we say that the swap is {\em exact}.
\item{}$\Delta a \leftarrow p.\mbox{\bf swapBtoA}(\Delta b)$, where $\Delta b$ is a positive real number. A negative real number $\Delta a$ is given as output. If the current state of the pool is $(a,b)$, the following inequality must be satisfied:
  \begin{equation}
    \label{alsoincrease}
    (a + \Delta a)(b + \Delta b) \geq ab.
  \end{equation}
  As a result of the operation, the pool changes state from $(a,b)$ to $(a + \Delta a, b + \Delta b)$ with $z$ unchanged. This models a swapper using the pool to exchange $\Delta b$ units of asset B to $-\Delta a$ units of asset A. 
\item{}$t \leftarrow p.\mbox{\bf provide}(\Delta a, \Delta b)$, where $\Delta a, \Delta b$ are positive real numbers so that there exists an $\epsilon > 0$ so that $\Delta a = \epsilon a$ and $\Delta b = \epsilon b$, where $(a,b)$ is the current asset amount pair of the pool. As a result of the operation, the pool changes state from $(a,b), z$ to $(a',b')=((1+\epsilon)a, (1+\epsilon)b), z'=(1+\epsilon)z$ and a liquidity token portion $t$ of amount $\epsilon z$ is given as output. This models a liquidity provider adding liquidity to the pool, in the ``correct proportions''.  
\item{}$(a^*, b^*) \leftarrow p.\mbox{\bf reclaim}(t)$, where $t$ is a liquidity token portion that is an output of a previous {\bf provide} operation or of the {\bf init} operation, and not previously given as input to a {\bf reclaim} operation. As a result of the operation, the pool changes state from $(a,b), z$ to $(a',b')=(a - a^*, b - b^*), z'=z-m$, where $m$ is the amount of the portion $t$, $a^* = am/z$ and $b^* = bm/z$. {\em We only allow this operation if $m < z$.} This models a liquidity provider returning his liquidity token portion and getting assets in return ($a^*$ units of A and $b^*$ units of B), while the liquidity token portion is burned. (To keep the model simple, we refrain from modeling the possibility of splitting the portion in subportions and burning the subportions at various points in time.)
\end{itemize}

An important property of liquidity pools which for the classical case follows directly from the construction (in particular, the restriction for the {\bf reclaim} operation), but which for pools we consider later is more subtle, is {\em asset positivity}.
\begin{fact}\label{asset-positivity} {\bf Asset positivity.}
  At any point in time, the state of a classical liquidity pool satisfies $a > 0$ and $b > 0$. 
\end{fact}
Asset positivity reflects the fact that the pool state variables $a,b$ model actual assets held by the pool (e.g., an amount of Bitcoin and an amount of Ethereum) and we want to ensure that the pool is not in debt. Thus, asset positivity implies that we are actually able to send the promised asset amounts to swappers when locks are executed and the correct amount of assets to liquidity providers doing a {\bf reclaim}.
  
A {\em trace} of a classical liquidity pool is a sequence of operations on the pool, starting with an {\bf init} operation and satisfying the constraints described above. The following is well-known and is a crucial property of a classical liquidity pool:
\begin{theorem}\label{impermanence}
  {\bf Product preservation. }  Consider an $(a^*, b^*) \leftarrow p$.{\bf reclaim}$(t)$ operation for some liquidity token portion $t$ in a trace of a classical liquidity pool $p$.
  Per the requirements of the definition of {\bf reclaim}, let $\Delta a, \Delta b$ be values provided as arguments in the earlier $t \leftarrow p.\mbox{\bf provide}(\Delta a, \Delta b)$ operation or $t \leftarrow p.\mbox{\bf init}(\Delta a, \Delta b)$ operation in the trace, i.e., the operation matching the {\bf reclaim} operation. Then,
  \begin{equation}
    a^*b^* \geq \Delta a \Delta b.
   \end{equation}
\end{theorem}
Informally stated, this means that if a liquidity provider adds $\Delta a$ units of asset A and $\Delta b$ units of asset B, receives liquidity tokens in return and later burns those liquidity token to receive $a^*$ units of asset A and $b^*$ units of asset B, then $a^*b^* \geq \Delta a \Delta b$.

Product preservation is a formalization of (a strengthening of) the well-known slogan of {\em impermanent loss} for liquidity providers. The impermanent loss slogan is not only well-known but also well known to be somewhat misleading. Les us recall the narrative behind the slogan and see what the connection is to the formal property of product preservation. Let us assume that a liquidity provider provides assets $\Delta a, \Delta b$ when the ratio $a/b$ between the pool assets is some value $p$, so we assume that $\Delta a/\Delta b = p$ as well. Note that $p$ can also be interpreted as the (marginal) price of asset B in units of asset A at that point in time. If the liquidity provider at some later arbitrary time $t$ decides to withdraw his assets, he may well suffer loss, but if he instead decides to  {\em wait} until the marginal price of asset B is once again $p$, then the withdrawn assets $a^*, b^*$ satisfies $a^*/b^* = p$ and product preservation then implies that we in fact have that $a^* \geq \Delta a$ and $b^* \geq \Delta b$. That is, the liquidity provider in fact has at least as much of both assets as he started out with, and the loss he {\em seemed} to have suffered at time $t$ was indeed impermanent. It is a nice narrative, but the issue is of course that {\em there is no guarantee that the price of asset $B$ will ever return to $p$.} Therefore, we prefer the more neutral terminology of product preservation.

Despite its connection to the somewhat dubious notion of impermanent loss, product preservation constitutes an important guarantee for liquidity providers. Note that a nice immediate corollary of product preservation is that it is not possible for a malicious third party to almost deplete a classical liquidity pool of its assets by some clever sequence of operations, as each provider of those assets is in possession of a guarantee that he can get the amount he originally provided of a least {\em one} of his two assets back (if one goes down, the other must come up).

The product preservation theorem can be proved by an easy induction, but we shall not do so here, as it is a very well-known property of classical liquidity pools. 
Rather, we will use it later to prove a similar statement for a lock-swap enabled liquidity pool, {\em the retroactive pool}, by reducing that statement to the statement for classical pools.

\section{Non-proportional-provide pools}
\label{npp}
In this section, we consider a natural extension of the notion of a classical liquidity pool, a {\em non-proportional-provide pool}. We are going to use this concept as a stepping stone towards the lock-swap enabled pool. This section also serves as an introduction to our general methodology; to prove properties of the extended pool concepts by reductions to properties of classical pools and in particular reductions to Theorem \ref{impermanence}.

A non-proportional-provide pool is defined as a classical liquidity pool, except that its {\bf provide} operation is defined as follows, allowing for arbitrary additions of liquidity to the pool, not necessarily in the ``correct proportions'':
\begin{itemize}
\item{}$t \leftarrow p.\mbox{\bf provide}(\Delta a, \Delta b)$, where $\Delta a, \Delta b$ are arbitrary non-negative numbers. As a result of the operation, the pool changes state from $(a,b), z$ to $(a',b')= (a + \Delta a, b + \Delta b), z' = z + m$, where
  \begin{equation}
    m = \epsilon z,
    \end{equation}
    with
  \begin{equation}
    \epsilon = \sqrt{\frac{(a + \Delta a)(b + \Delta b)}{ab}} - 1.
  \end{equation}
    and a liquidity token portion $t$ of amount $m = \epsilon z$ is given as output.
\end{itemize}
Asset positivity is immediate for non-proportional-provide pools. More interestingly, and perhaps somewhat surprisingly, the product preservation theorem is valid {\em unmodified} for non-proportional-provide pools. While not difficult to prove, we believe this fact is not as well-known as it should be and consider it of independent interest.
\begin{theorem}\label{impermanence2}
  {\bf Product preservation for non-proportional-provide pools. }Consider an $(a^*, b^*) \leftarrow p$.{\bf reclaim}$(t)$ operation for some liquidity token portion $t$ in a
  trace of a non-proportional-provide pool $p$. Per the requirements of the definition of {\bf reclaim}, let $\Delta a, \Delta b$ be values provided as arguments in the earlier $t \leftarrow p.\mbox{\bf provide}(\Delta a, \Delta b)$ operation or $t \leftarrow p.\mbox{\bf init}(\Delta a, \Delta b)$ operation in the trace, i.e., the operation matching the {\bf reclaim} operation. Then,
  \begin{equation}
    a^*b^* \geq \Delta a \Delta b.
   \end{equation}
\end{theorem}
\begin{proof}
  The proof is a reduction to the case of classical pools: Given a trace of a non-proportional-provide pool, we construct a {\em corresponding} trace of a classical pool and appeal to Theorem \ref{impermanence}. To construct this classical trace, we simply copy all operations except each non-proportional {\bf provide} which much be handled differently. To deal with such a non-proportional {\bf provide}$(\Delta a, \Delta b)$, we first observe that if the proportions are in fact correct, i.e. $\Delta a/\Delta b = a/b$, where $(a,b)$ is the pool state, the formulas given in the definition of {\bf provide} matches the formula given in the classical case. In that case, we can just copy the operation to the corresponding trace.
  Observe next that it is possible to add arbitrary amounts of assets $(\Delta a, \Delta b)$ with ``wrong proportions'' to a classical liquidity pool by
  \begin{itemize}
  \item{}first doing a single exact swap of part of our holding of one of the two assets (the one for which we have too much) to the other asset (the one for which we have too little), making our holding into a pair $(a_1,b_1)$ with the correct proportions for addition to the pool, and
  \item{}subsequently adding $(a_1,b_1)$ to the pool with a classical {\bf provide} operation.
  \end{itemize}
  The crucial property that we shall need is the following:
  \begin{equation}\label{increase2}
    a_1b_1 \geq \Delta a \Delta b.
  \end{equation}
  We give an algebraic proof of (\ref{increase2}). By assumption, we have assets $(\Delta a, \Delta b)$ in ``wrong proportions'', that is, $\Delta a/\Delta b \not = a/b$, where $(a,b)$ is the pool state. We can assume without loss of generality that $\Delta a/\Delta b > a/b$. In that case, we swap an amount $x$ of asset A into an amount $y$ of asset B (so $a_1 = \Delta a - x$ and $b_1 = \Delta b + y$), with the following two equations satisfied:
  \begin{equation}\label{correctswap}
    (a + x)(b - y) = a b,
  \end{equation}
  \begin{equation}\label{proportional}
    \frac{\Delta a - x}{\Delta b + y} = \frac{a + x}{b - y}
  \end{equation}
  where (\ref{correctswap}) is an instantiation of (\ref{increase}), taking into account that the swap is exact, and (\ref{proportional}) expresses that we choose $x$ to make the new pair $(\Delta a - x, \Delta b + y)$ the correct proportions for addition to a classical pool, taking into account that the swap also changes the proportions of the pool. Equations (\ref{correctswap}-\ref{proportional}) are two quadratic equations in two unknowns which we can solve symbolically, yielding:
  \begin{equation}\label{eqx}
    x = - a + \sqrt{a b (\frac{a + \Delta a}{b + \Delta b})},
  \end{equation}
  \begin{equation}\label{eqy}
    y = b - \sqrt{a b (\frac{b + \Delta b}{a + \Delta a})}.
  \end{equation}
  We now want to establish (\ref{increase2}), i.e., that $(\Delta a -x)(\Delta b + y) \geq \Delta a \Delta b$. Plugging in the values of $x, y$ from (\ref{eqx}-\ref{eqy}), we get
  \begin{eqnarray*}
    & & (\Delta a -x)(\Delta b + y) \\
    & = & (\Delta a + a - \sqrt{a b (\frac{a + \Delta a}{b + \Delta b})}))(\Delta b + b - \sqrt{a b (\frac{b + \Delta b}{a + \Delta a})})\\
    & = & \Delta a \Delta b + 2ab + a \Delta b + b \Delta a - (a + \Delta a) \sqrt{a b (\frac{b + \Delta b}{a + \Delta a})} - (b + \Delta b) \sqrt{a b (\frac{a + \Delta a}{b + \Delta b})}\\
    & = & \Delta a \Delta b + 2ab + a \Delta b + b \Delta a - 2 \sqrt{a b (a + \Delta a)(b + \Delta b)}
  \end{eqnarray*}
  That is, we want to show
  \begin{equation}
    \Delta a \Delta b + 2ab + a \Delta b + b \Delta a - 2 \sqrt{a b (a + \Delta a)(b + \Delta b)} \geq \Delta a \Delta b
  \end{equation}
  or, equivalently
\begin{equation}
    2ab + a \Delta b + b \Delta a  \geq  2 \sqrt{a b (a + \Delta a)(b + \Delta b)}
\end{equation}
or, equivalently
\begin{equation}
    (2ab + a \Delta b + b \Delta a)^2  \geq  4 a b (a + \Delta a)(b + \Delta b)
\end{equation}
or, equivalently
\begin{equation}
  (b\Delta a - a\Delta b)^2 \geq 0
\end{equation}
which is true. This establishes (\ref{increase2}).
After the swap, we provide assets $(a_1,b_1) = (\Delta a - x, \Delta b + y)$ to the classical pool, this being possible because (\ref{proportional}) establishes that there is an $\epsilon > 0$ so that
$\Delta a - x = \epsilon (a + x)$ and $\Delta b + y = \epsilon (b - y)$. By the definition of the classical {\bf provide} operation, the pool changes state to $(a + \Delta a, b + \Delta b) = (1+\epsilon)(a + x, b - y)$ and
a liquidity token portion $t$ of amount $\epsilon z$ is given as output. Thus, the pool state of the constructed corresponding classical trace matches the pool state of the given non-proportional-provide trace. This completes our handling of the case of a non-proportional-provide operation.

We are now ready to prove the statement of the theorem.

Given a ``matching pair'' $(a^*, b^*) \leftarrow p$.{\bf reclaim}$(t)$ and $t \leftarrow p.\mbox{\bf init}(\Delta a, \Delta b)$ of operations, we have $a^*b^* \geq \Delta a \Delta b$ as this is the case for the corresponding classical trace, by Theorem \ref{impermanence}.

Given a ``matching pair'' $(a^*, b^*) \leftarrow p$.{\bf reclaim}$(t)$ and $t \leftarrow p.\mbox{\bf provide}(\Delta a, \Delta b)$ of operations, let $(a_1,b_1)$ be the result of the swap used when simulating the {\bf provide} operation in the corresponding classical trade, as described above. By (\ref{increase2}), we have $a_1 b_1 \geq \Delta a \Delta b$ and by Theorem \ref{impermanence}, we have
$a^*b^* \geq a_1b_1$. That is, $a^*b^* \geq \Delta a \Delta b$. This completes the proof.
  \end{proof}
\section{The lock-swap enabled liquidity pool as an abstract data type}
\label{lockswap}
We define a {\em lock} to be a pair $(\Delta a, \Delta b)$, where exactly one value in the pair is positive and exactly one is negative. Intuitively, a lock represents an {\em option} to change the content of the pool by $\Delta a, \Delta b$. As for the case of liquidity token portions, we consider two locks different, even if the amounts in the two pairs are the same (as we need to keep track of particular locks in the statements and proofs).

The state of a  {\em lock-swap enabled liquidity pool} includes at any point in time an asset amount pair $(a,b)$ of two positive real numbers $a > 0$ and $b > 0$ and another positive real number $z$, with the same intuition behind these variables as in previous sections. In particular $a,b$ are to be thought of as amounts of assets (e.g. Bitcoin, Ethereum) literally held by the pool and $z$ as the number of liquidity tokens currently distributed among liquidity providers.

The lock swap enabled pool is accessed and manipulated using the following operations:

\begin{itemize}
  \item{} $t \leftarrow p.\mbox{\bf init}(\Delta a, \Delta b)$, where $\Delta a, \Delta b$ are positive real numbers. This initializes the pool to $(a,b) = (\Delta a, \Delta b)$ and $z = 1$ and returns a liquidity token portion $t$ of amount 1. This operation is only permitted once and must be the first operation performed.
\item{}$L \leftarrow p.\mbox{\bf lockSwapAtoB}(\Delta a)$, where $\Delta a$ is a positive real number, and $L$ is a lock returned by the operation whose $\Delta a$ part matches the $\Delta a$ given as argument to the {\bf lockSwapAtoB} operation (so the $\Delta b$-part of the lock is negative). {\em It is a requirement that any lock received in this way is eventually either executed or canceled} (for definitions of execution and cancelation, see below). A lock that has been received in a lockSwap operation but not yet executed or canceled is called {\em active} or {\em unresolved}.
\item{}$L \leftarrow p.\mbox{\bf lockSwapBtoA}(\Delta b)$, where $\Delta b$ is a positive real number, and $L$ is a lock whose $\Delta b$ part matches the $\Delta b$ given as argument to the {\bf lockSwapBtoA} operation (so the $\Delta a$-part of the lock is negative). Similar conditions as for the {\bf lockSwapAtoB} operation applies.
\item{}{\bf execute}$(L)$, where $L$ is a lock with content $(\Delta a, \Delta b)$ that was previously received as output of a {\bf lockSwapAtoB} or {\bf lockSwapBtoA} operation and not previously given as input to an {\bf execute} or {\bf cancel} operation. The operation changes the pool state from $(a,b)$ to $(a',b') = (a + \Delta a, b + \Delta b)$, while leaving $z$ unchanged. Intuitively, but not explicitly part of the model, corresponding asset amounts are sent/received to/from the swapper. 
\item{}{\bf cancel}$(L)$, where $L$ is a lock previously received as output of a {\bf lockSwapAtoB} or {\bf lockSwapBtoA} operation and not previously given as input to an {\bf execute} or {\bf cancel} operation. The operation does absolutely nothing, except preventing the lock from ever being executed, as per the definition of the {\bf execute} operation.
\item{}$t \leftarrow p.\mbox{\bf provide}(\Delta a, \Delta b)$, where $\Delta a, \Delta b$ are non-negative numbers. This changes the pool state from $(a,b)$ and $z$ to $(a',b') = (a + \Delta a, b + \Delta b)$ and $z' = z + m$ and a liquidity token portion $t$ of amount $m$ is returned. For technical reasons we discuss later, we allow the liquidity token portion $t$ to be available for later operations (intuitively, ``sent to the provider'') {\em only when all locks granted at the time of the {\bf provide} operation have been either executed or canceled.} Until this happens, we say that the operation is unresolved. 
\item{}$(a^*, b^*) \leftarrow p.\mbox{\bf reclaim}(t)$, where $t$ is a liquidity token portion that is an output of a previous {\bf provide} operation or of the {\bf init} operation, and not previously given as an argument to a  {\bf reclaim} operation. As a result of the operation, the $z$-part of the pool state changes to $z-m$, where $m$ is the amount of the liquidity token portion. {\em We only alllow this operation if $m < z$.} It also changes the pool state from $(a,b)$ to $(a - a^*, b - b^*)$ with
  $a^*, b^*\geq 0$. This models a liquidity provider returning his liquidity token portion and getting assets $(a^*, b^*)$ in return. For technical reasons and similar to the case of {\bf provide}, we allow this state change (subtraction of $a^*, b^*$ to the assets held in the pool and the asset amounts  $(a^*, b^*)$ being sent to the provider) to take place {\em only when all locks
  granted at the time of the {\bf reclaim} operation have been either executed or canceled.} Until this happens, we say that the operation is unresolved.
\end{itemize}
Note that we do not include the ``classical'' swap operations {\bf swapAtoB} and {\bf swapBtoA} in the definition, as these can be simulated by aquiring a lock and immediately executing it. 
Note also that the definition is more abstract than the previous ones and that one can imagine various implementations. However, to be a correct implementation, we want the usual central properties of liquidity pools to be satisfied:
\begin{itemize}
\item{}{\bf Asset positivity}: At all times, $a> 0$ and $b > 0$.
\item{}{\bf Product preservation: }Consider an $(a^*, b^*) \leftarrow p$.{\bf reclaim}$(t)$ operation for some liquidity token portion $t$ in a trace of a lock-swap enabled pool $p$.
 Per the requirements in the definition of {\bf reclaim}, let $\Delta a, \Delta b$ be values provided as arguments in the earlier $t \leftarrow p.\mbox{\bf provide}(\Delta a, \Delta b)$ operation or $t \leftarrow p.\mbox{\bf init}(\Delta a, \Delta b)$ operation in the trace, i.e., the operation matching the {\bf reclaim} operation. Then,
  \begin{equation}
    a^*b^* \geq \Delta a \Delta b.
   \end{equation}
\end{itemize}
These properties are of course analogous to the corresponding proven properties of classical liquidity pools and non-proportional-provide pools, but are now viewed as correctness criteria for an implementation of the lock-swap enabled pool, viewed as an abstract data type.
In addition to these formal requirements, we may have some informal ones: The quotes given in the locks should reflect ``reasonable'' prices and liquidity provided should be ``put to work'' in the pool immediately. But even without these informal requirements, the construction of lock-swap enabled pools is not trivial. In the next section, we present our main concrete construction.

\section{The retroactive pool, a particular lock-swap enabled pool}
\label{retroactive}
A {\em retroactive pool} $p$ is our main construction of an implementation of a lock-swap enabled liquidity pool. It is defined as follows.
The pool state is given by:
\begin{itemize}
\item{}The variables $a,b,z$ with same meaning as in the previous cases. In particular, these variables model amounts of actual holdings of assets by various parties (the pool and the liquidity providers).
\item{}A set $T$ of traces of non-proportional-provide pools. We call these pools the {\em virtual pools}. At any point in time, there is one element of $T$ for each way of resolving (executing or canceling) each of the locks active in $p$ at the time.
Thus, if there at some point in time are $k$ active locks in $p$, we have $|T| = 2^k$. Each of these non-proportional-provide traces, defined by a particular combination $\rho$ of executing or canceling each of the active locks, is derived from the trace of the retroactive pool $p$ in the following way:
\begin{itemize}
\item{}The {\bf init} operation is the same as the {\bf init} operation of the retroactive pool $p$.
\item{}Each {\bf provide} and {\bf reclaim} operation in the trace of $p$ is replaced with its non-proportional-provide counterpart (with the same inputs and correct outputs as defined for a non-proportional-provide pool).
\item{}Each {\bf execute} and {\bf cancel} operation in the trace of $p$ is removed.
\item{}Each {\bf lockSwapAtoB} and {\bf lockSwapBtoA} operation in the trace of $p$ that is later canceled in the trace of $p$ is removed.
\item{}Each {\bf lockSwapAtoB} and {\bf lockSwapBtoA} operation in the trace of $p$ that is later executed in the trace of $p$ is replaced with a {\bf swapAtoB} or {\bf swapBtoA} operation with the same parameters $\Delta a, \Delta b$.
\item{}Each remaining (unresolved) {\bf lockSwapAtoB} operation and {\bf lockSwapBtoA} operation in the trace of $p$ that is to be executed according to $\rho$ is replaced with a {\bf swapAtoB} or {\bf swapBtoA} operation with the same parameters $\Delta a, \Delta b$. Each {\bf lockSwapAtoB} and {\bf lockSwapBtoA} operation that is to be canceled according to  $\rho$ is removed.
\end{itemize}
\end{itemize}
Let us see how to maintain this pool state.
\begin{itemize}
\item{}To initialize the retroactive pool $p$, we let $T$ be a singleton set containing a non-proportional-provide trace with the same initialization operation.
\item{}To {\bf execute} a lock $L$, we remove from $T$ all traces corresponding to combinations where $L$ was canceled.
\item{}To {\bf cancel} a lock $L$, we remove from $T$ all traces corresponding to combinations where $L$ was executed.
\item{}To do a {\bf provide} or {\bf reclaim} operation, we extend each trace in $T$ with its non-proportional-provide counterpart (with the same inputs and correct outputs as defined for a non-proportional-provide pool).
\item{}To do a {\bf lockSwapAtoB}$(\Delta a)$ operation, we extend each of the non-proportional-provide traces in $T$ with a $\Delta b \leftarrow {\bf swapAtoB}(\Delta a)$ operation, in some correct way (exact or with some positive fee). Each of those newly extended traces has some negative $\Delta b$ value returned by the last operation. Let $\beta$ be the {\em largest} of those values (i.e. the negation of the {\em smallest} negated value). Now replace the $\Delta b$-value in the lock in the final operation in each of the traces with $\beta$. As (\ref{increase}) is an inequality, the traces are still correct non-proportional-provide traces (but possibly with larger fees in the last operation than before this final modification). The lock $(\Delta a, \Delta b)$ we give as output to the lock-swap operation has $\Delta b = \beta$.
\item{}Similarly, to do a {\bf lockSwapBtoA}$(\Delta b)$ operation, we extend each of the non-proportional-provide traces in $T$ with a $\Delta a \leftarrow {\bf swapAtoB}(\Delta b)$ operation, in some correct way (exact or with some positive fee). Each of those newly extended traces has some negative $\Delta a$ value returned by the last operation. Let $\alpha$ be the {\em largest} of those values (i.e. the negation of the {\em smallest} negated value). Now replace the $\Delta a$-value in the lock in the final operation in each of the traces with that value, $\alpha$. As (\ref{alsoincrease}) is an inequality, the traces are still correct non-proportional-provide traces (but possibly with larger fees in the last operation than before this final modification). The lock $(\Delta a, \Delta b)$ we give as output to the lock-swap operation has $\Delta a = \alpha$.
\end{itemize}
As each lock granted is eventually either canceled or executed, the evolving set $T$ has the property that for any natural number $n$, at some point in time, all traces in $T$ will agree on the first $n$ items. By increasing $n$, this defines an arbitrarily long trace of a non-proportional-provide pool. This trace we call the {\em retroactive trace} of $p$. Note that {\bf provide} and {\bf reclaim} operations become part of the retroactive trace when all locks active at the time of their execution are resolved. When a {\bf provide} operation becomes part of the retroactive trace, we finalize its execution by modifying $z$ as stipulated by the operation and sending the correct amount of liquidity tokens to the provider. When a {\bf reclaim} operation becomes part of the retroactive trade, we finalize its execution by modifying $z$ as stipulated by the operation and sending the correct amount of assets to the provider.

\begin{theorem}
  The retroactive pool is a correct implementation of a lock-swap enabled pool. That is, it satifies asset positivity and product preservation.
\end{theorem}
\begin{proof}

Perhaps surprisingly, it is easier to establish product preservation, so we do this first. We simply observe that the product preservation condition, as stated, is in fact a property of the retroactive trace. In other words, the retroactive pool satisfies product preservation if and only if its retroactive trace does. But the retroactive trace is a trace of a non-proportional-provide pool, which by Theorem \ref{impermanence2} satisfies product preservation, so we are done.

Asset positivity is more subtle, as this is {\em not} a property of the retroactive trace, so we need to do more work. Assume, to the contrary, that a retroactive pool $p$ does not satisfy asset positivity. Then, at some point in a trace of $p$, either $a$ or $b$ is reduced to a value at or below 0. Let us assume, without loss of generality, that it is in fact $a$ that is reduced to a non-positive number. As {\bf provide} operations increase the asset amounts and the lock-swap operations and the {\bf cancel} operation leave them unchanged, this could only happen in two ways:
  \begin{enumerate}
  \item{}During the final, delayed execution of a {\bf reclaim} operation.
  \item{}During an {\bf execute} operation.
   \end{enumerate}
  We can rule out the first case as follows. The final, delayed execution of a {\bf reclaim} operation happens when the operation becomes part of the retroactive trace. But then, we are in good shape: Again, the retroactive trace is a trace of a non-proportional-provide pool, and a non-proportional-provide pool satisfies asset positivity. Hence, $a$ cannot become non-positive by the operation. This leaves the case of an {\bf execute} operation which is more subtle. Consider the point in time where an {\bf execute}-operation supposedly reduces $a$ to a non-positive number and let $T$ the the set of traces at that point in time. Consider the trace in $T$ corresponding to the virtual pool $\tilde t$ in which all active locks are canceled. We claim that that trace also must fail to satisfy asset positivity.
  The claim follows from the fact that the {\em multi-set} of actual modifications to the assets in the form of additions of $(\Delta a, \Delta b)$ pairs
  made by operations in the trace of the retroactive pool $p$ is the same as the {\em multi-set} of such actual modifications to the assets in the virtual pool $\tilde p$, except that the trace of $\tilde p$ may contain some modifications $(-a^*, -b^*)$ from {\bf reclaim} operations, which will only reduce the value of $a$. As addition is commutative, we have that the current value of $a$ in the pool $p$ is at most the final value of $a$ in the trace of the pool $\tilde p$, i.e., they are both negative. But this is a contradiction, as the trace of $\tilde p$ is the trace of a non-proportional-provide pool and thus satisfies asset positivity.
 \end{proof}

\section{A python implementation of the retroactive pool}
\label{python}
We have made a python3 prototype of the retroactive pool. The key to this (fairly) practical prototype is to be mindful about how to implement the set of traces $T$ of the virtual pools.
We do it as follows: We do not explicitly represent the traces. Rather, we explicitly represent the trace of the retroactive pool itself and use the fact that the set of traces $T$ can be derived from this by enumerating all combinations of cancelations and executions of active locks. This is conveniently represented in python as a generator for the set $T$. Also, rather than explicitly representing the {\em entire} trace of the retroactive pool, we represent a compressed version, containing all necessary information. The trace is compressed in two ways: We do not represent events earlier that the earliest unresolved lock (that is, we do not represent the retroactive trace), and we do not represent each and every resolved lock; resolved locks are only relevant for what they added to and removed from the pool, so adjacent resolved locks are combined into one, again using commutativity of addition.

More formally, the compressed trace consists of:
\begin{enumerate}
\item{}An initial asset allocation $a_0, b_0$ which is really the pool allocation resulting from the sequence of operations in the retroactive trace.
\item{}An ordered sequence of pool operations where each entry is either
  \begin{enumerate}
   \item{}A description of an unresolved lock, with various associated information
   \item{}A description of an unresolved {\bf provide}
      (liquidity that was provided to the pool, but for which unresolved locks
      prevent us sending pool tokens to the provider just yet)
   \item{}A description of an unresolved {\bf reclaim}
      (liquidity that was reclaimed from the pool, but for which unresolved locks
      prevent us returning assets to the provider just yet)
   \item{}A description of a resolved lock, or multi-lock
      (a multi-lock being several resolved locks combined)
  \end{enumerate}
  \end{enumerate}
with the invariants that
\begin{enumerate}
\item{}either the sequence is empty or its first entry is of type (a)
\item{}there are no two entries of type (d) in a row
\end{enumerate}
The invariants imply that the representation size of the pool state is proportional to the current number of unresolved operations (locks, {\bf provide}s, {\bf reclaim}s).

The resulting python prototype can be found in Appendix A. Running it (using ``python3 retroactive.py'') will test it on a randomly generated sequence of operations, with the correctness criteria of asset positivity and product preservation being continuously asserted. As the complexity of carrying out the computation in a lock-swap operation is lower bounded by the size of $T$ and hence exponential in the number of active locks, the number of simultaneous active locks is kept smaller than or equal to 17.

\section{The retroactive pool with restricted liquidity changes}
\label{restricted}
In this section, we consider restrictions to the operations that will enable us to avoid the unfortunate exponential (in the number of unresolved locks) time complexity for carrying out an operation of the retroactive pool. It is convenient to use the concrete code of the prototype to discuss these restrictions and how they work. In Figure \ref{fig-gen} is the python generator representing the set of final allocations $(a_\tau,b_\tau)$ of the traces $\tau \in T$ (with $T$ as in the description of the retroactive pool).

\begin{figure}
\begin{python}
    def all_possible_allocations(self):
        k = self.no_unresolved_locks()
        cases = raise_power(2, k)
        for rho in range(0, cases):
            my_rho = rho
            a = self.a0
            b = self.b0
            no_tokens = self.no_liquiditytokens_minted
            for e in self.events:
                if isinstance(e, UnresolvedLock):
                    if is_even(my_rho):
                        a += e.lock.delta_a
                        b += e.lock.delta_b
                    my_rho = my_rho // 2
                elif isinstance(e, ResolvedLocks):
                    a += e.delta_a
                    b += e.delta_b
                elif isinstance(e, UnresolvedProvide):
                    new_a = a + e.delta_a
                    new_b = b + e.delta_b
                    delta = new_a * new_b / (a * b) - 1
                    epsilon = math.sqrt(1 + delta) - 1
                    no_tokens += epsilon * no_tokens
                    a = new_a
                    b = new_b
                else:
                    assert isinstance(e, UnresolvedReclaim)
                    burn_tokens = e.liquiditytokens.no_tokens
                    pool_fraction = burn_tokens / no_tokens
                    no_tokens -= burn_tokens
                    a *= 1 - pool_fraction
                    b *= 1 - pool_fraction
            yield a, b
\end{python}
\caption{Python generator implicitly representing the set of traces $T$ maintained in the pool state of the retroactive pool  }\label{fig-gen}
\end{figure}

Recall from the descripion of the retroactive pool that the computation that causes the exponential time complexity is the following:
When lock-swapping $\Delta a$ units of asset A to asset B, we must, among final allocations $(a_\tau, b_\tau), \tau \in T$, find the one that gives the {\em lowest} value of $-\Delta b$ if we do a swap (and similarly, when swapping from B to A). The following easy-to-prove theorem identifies the relevant trace directly in some natural special cases, eliminating the need for an exponential search.  
\begin{theorem}\label{only-two-virtual-pools}
  Consider a retroactive pool in some current state. Suppose at least one of the following is true:
  \begin{itemize}
\item{}There are no unresolved {\bf reclaim}s in the current state.
\item{}There are no unresolved {\bf provide}s in the current state.
  \end{itemize}
  Then, the trace $\tau \in T$ that gives the lowest value of the output $-\Delta b$ for an exact swap with input $\Delta a$ to the pool with allocation $(a_\tau, b_\tau)$, is the trace $\tau$ corresponding to the virtual pool that cancels all unresolved locks for swaps from B to A and executes all unresolved locks for swaps from A to B. The analogous statement holds for swaps in the other direction.
\end{theorem}
\begin{proof}
  First consider the case of no unresolved {\bf reclaim}s. In that case, the computation of liquidity tokens for the case of unresolved {\bf reclaim}s becomes irrelevant in the code of Figure \ref{fig-gen} and we can simplify the code to the code of Figure \ref{fig-gen2}.
  \begin{figure}
\begin{python}
    def all_possible_allocations(self):
        k = self.no_unresolved_locks()
        cases = raise_power(2, k)
        for rho in range(0, cases):
            my_rho = rho
            a = self.a0
            b = self.b0
            no_tokens = self.no_liquiditytokens_minted
            for e in self.events:
                if isinstance(e, UnresolvedLock):
                    if is_even(my_rho):
                        a += e.lock.delta_a
                        b += e.lock.delta_b
                    my_rho = my_rho // 2
                else:
                assert isinstance(e, ResolvedLocks)
                    or isinstance(e, UnresolvedProvide):
                    a += e.delta_a
                    b += e.delta_b
            yield a, b
\end{python}
\caption{Python generator implicitly representing the set of traces $T$, no unresolved {\bf reclaim}s }\label{fig-gen2}
  \end{figure}Then, consider any particular unresolved lock $L$ containing $(\alpha, \beta)$, where exactly one of $\alpha$ and $\beta$ is positive and the other negative, and consider the decisions about whether to execute or cancel all other locks as already having been made. Suppose canceling the lock $L$ will result in yielding values $a = a^*, b = b^*$. As addition is commutative, executing the lock will then result in the yielding values $a = a^* + \alpha, b = b^* + \beta$. Consider doing the same swap from A to B in the two corresponding pools. It is immediate that for the case of negative $\beta$ and any input $\Delta a$, we get the lowest value of $-\Delta b$ when swapping in the pool $a = a^* + \alpha, b = b^* + \beta$, while for the case of negative $\alpha$, we get the lowest value of $-\Delta b$ when swapping in the pool $a = a^*, b = b^*$. The statement of the theorem follows.

Next, consider the case of no unresolved {\bf provide}s. Now, the computation of liquidity tokens for the case of unresolved {\bf provide}s becomes irrelevant in the code of Figure \ref{fig-gen} and we can simplify the code to the code of Figure \ref{fig-gen3}.
\begin{figure}
\begin{python}
    def all_possible_allocations(self):
        k = self.no_unresolved_locks()
        cases = raise_power(2, k)
        for rho in range(0, cases):
            my_rho = rho
            a = self.a0
            b = self.b0
            no_tokens = self.no_liquiditytokens_minted
            for e in self.events:
                if isinstance(e, UnresolvedLock):
                    if is_even(my_rho):
                        a += e.lock.delta_a
                        b += e.lock.delta_b
                    my_rho = my_rho // 2
                elif isinstance(e, ResolvedLocks):
                    a += e.delta_a
                    b += e.delta_b
                else:
                    assert isinstance(e, UnresolvedReclaim)
                    burn_tokens = e.liquiditytokens.no_tokens
                    pool_fraction = burn_tokens / no_tokens
                    no_tokens -= burn_tokens
                    a *= 1 - pool_fraction
                    b *= 1 - pool_fraction
            yield a, b
\end{python}
\caption{Python generator implicitly representing the set of traces $T$, no unresolved {\bf provide}s }\label{fig-gen3}
\end{figure}
Then, consider any particular unresolved lock $L$ containing $(\alpha, \beta)$, where exactly one of $\alpha$ and $\beta$ is positive and the other negative, and consider the decisions about whether to execute or cancel all other locks as already having been made. Suppose canceling the lock $L$ will result in yielding values $a = a^*, b = b^*$. As addition is commutative, executing the lock will then result in the yielding values $a = a^* + \lambda \alpha, b = b^* + \lambda \beta$, where $\lambda$ is a scaling factor in the interval $(0,1)$ that depends on the unresolved {\bf reclaim}s appearing in the trace after the lock. Consider doing the same swap from A to B in the two corresponding pools. It is again immediate that for the case of negative $\beta$ and any input $\Delta a$, we get the lowest value of $-\Delta b$ when swapping in the pool $a = a^* + \lambda \alpha, b = b^* + \lambda \beta$, while for the case of negative $\alpha$, we get the lowest value of $-\Delta b$ when swapping in the pool $a = a^*, b = b^*$. The statement of the theorem follows.

The proof of the case of swaps from B to A is analogous.
\end{proof}

This means that if we make sure that at least one of the two cases of Theorem \ref{only-two-virtual-pools} holds at all times, we just need {\em two} virtual pools (one for each swap direction) rather than exponentially many.
What about the general case? Is it possible to restrict the number of virtual pools we look at to, say, two or another small constant without making restrictions on the applicability of {\bf provide} and {\bf reclaim}? This may very well be the case, but it is not as easily established as in the restricted cases: It is in general {\em not} the case that the allocation $(a_\tau, b_\tau)$ that gives the minimum $-\Delta b$ value is given by the trace of that virtual pool which cancels all active locks for swaps from B to A and executes all active locks for swaps from A to B. This is established by the following somewhat surprising counterexample which manipulates the number of minted but not burnt liquidity tokens in a subtle way. We present the example, but leave it as a fun exercise for the reader to understand and appreciate it.

Consider the following sequence of operations on a retroactive pool:
\begin{itemize}
\item{}$t_1 \leftarrow p.${\bf init}(1,1)
\item{}$L = p.${\bf lockSwapAtoB}(1000)
\item{}$t_2 \leftarrow p.${\bf provide}(172,0)
\item{}$(a^*, b^*) \leftarrow p.\mbox{\bf reclaim}(t_1)$
\item{}$t_3 \leftarrow p.{\bf provide}(0,10000)$
\end{itemize}
Since there is only one active lock, there are only two virtual pools. The final allocation in the one canceling the lock $L$ is $$(a,b) = (159.84705\ldots, 10000.92397\ldots).$$ This means that the marginal price of a unit of asset B in units of A is $0.01598 \ldots$. That is, for small $\Delta a$, we will have $-\Delta b \approx \Delta a/0.01598$, when swapping $\Delta a$ units of A to $-\Delta b$ units of B.

On the other hand, the final allocation in the virtual pool executing the lock $L$ with an exact swap is $$(a,b) = (89.40736\ldots, 10000.00007\ldots).$$ This means that the marginal price of a unit of asset B in units of A is $0.00894\ldots$. That is, for small $\Delta a$, we will have $-\Delta b \approx \Delta a/0.00894$, when swapping $\Delta a$ units of A to $-\Delta b$ units of B.

In other words, we get a higher marginal price of asset B in the end, and hence a smaller amount of B in return for a given, small amount of A  if we ``make a queen sacrifice'' and cancel the lock, i.e. refrain from ``locally'' making asset B as expensive as possible by executing the lock (the number 172 in the example was chosen to maximize this price difference). This example complicates matters. We do not know how complicated matters can get! (We also do not know if there is a way for a malicious party to exploit this issue if it is simply ignored). So we ask:

{\bf Open problem: }In the general case, assuming exact swaps in the virtual pools (to make the problem well-specified), can the minimum $-\Delta b$ be computed in polynomial (or linear) time? Alternatively, is it NP-hard? If it is NP-hard, can we find an efficient approximation algorithm or an exact algorithm with milder exponential complexity than the obvious one?

We return our attention to the restricted case described in Theorem \ref{only-two-virtual-pools} and the practical implications of that theorem.
We can ensure that one of the two cases hold by simply disallowing {\bf reclaim}s  whenever there are unresolved {\bf provide}s and vice versa. As {\bf provide}s and {\bf reclaim}s are much rarer operations than swaps in practice,  this may be an acceptable restriction. This reduces the time complexity of performing an operation on the pool to linear in the number of currently unresolved operations.

We can do even better in terms of time complexity if we simply disallow {\bf reclaim}s in the presence of unresolved locks. Then, the relevant set of traces is always generated by the code of Figure \ref{fig-gen2} and Theorem \ref{only-two-virtual-pools} furthermore implies that only two of these are relevant. Inspecting the code of Figure \ref{fig-gen2}, and also noticing that in absense of unresolved reclaims, the current actual assets held in the pool are given by the $a_0, b_0$ values together with the contributions of all executed locks and all unresolved {\bf provide} operations, we have:
\begin{corollary}
  Consider a retroactive pool in some current state. In particular, let $a,b$ be the current amounts of actual liquidity in the pool and let $T$ be the set of traces of virtual pools. Suppose that there are no unresolved {\bf reclaim}s in the current state. Let:
\begin{itemize}
\item{}$\Sigma^{\mbox{\rm{\tiny in}}}_a$
  denote the total incoming liquidity in unresolved locks from A to B (sum of $\Delta a$ values in those locks) and
\item{}$\Sigma^{\mbox{\rm{\tiny out}}}_b$
  denote total outgoing liquidity in unresolved locks from A to B (sum of $\Delta b$ values in those locks - this number is negative),
\end{itemize}
Then, the trace $\tau \in T$ that gives the lowest value of the output $-\Delta b$ for an exact swap with input $\Delta a$ to the virtual pool with allocation $(a_\tau, b_\tau)$, has:
\begin{equation}
  a_\tau = a + \Sigma^{\mbox{\rm{\tiny in}}}_a.
\end{equation}
\begin{equation}
  b_\tau = b + \Sigma^{\mbox{\rm{\tiny out}}}_b
\end{equation}
The analogous statement holds for swaps in the other direction.
\end{corollary}
We can keep track of the values $\Sigma^{\mbox{\rm{\tiny in}}}_a$ and $\Sigma^{\mbox{\rm{\tiny out}}}_b$ in constant time per operation on the pool, and if we do, the corollary implies that we can compute the relevant allocation $(a_\tau, b_\tau)$  in constant time.

\section{Discussion}
\label{discuss}
We have presented a new functionality for constant product liquidity pools, the {\em lock-swap}, and described an implementation of this functionality, the {\em retroactive pool}.

The lock-swap is a powerful tool for a trader. It is essentially a financial option, and options being valuable means that they are usually not handed out for free, indiscriminately. Also, the concrete implementation of the lock-swap in the retroactive pool will affect other traders operating on the pool in significant and predictable ways. In particular, recalling how the retroactive trace is formed, we see that if a lock-swap is eventually canceled, traders swapping in the pool in the same direction as the lock-swap, between the time of the lock-swap being granted and it being canceled, will effectively have paid a (possibly much) higher fee for their swap than the standard one, according to the retroactive trace.  
Similarly, if a lock-swap is eventually executed, traders swapping in the pool in the opposite direction as the lock-swap, between the time of the lock-swap being granted and it being executed, will effectively pay such higher fee. Thus, if the pool is active, with swaps in both directions, liquidity providers (who collect the fees) could benefit significantly from a lock-swap being granted to a traders. Other traders pay for this, as already mentioned.

This all suggests that one would probably want to regulate the use of lock-swaps. Furthermore, our construction requires that every lock-swap is eventually either executed or canceled, so some kind of regulation would have to enforce this  property anyway. Possibilities include:
\begin{itemize}
\item Making lock-swaps time-limited, automatically canceling them after a certain amount of time.
\item Putting a heavy fee on lock-swaps, with the fee increasing the longer the lock is held.
\item Only granting lock-swaps to special parties. This is a particularly attractive possibility. If such special party is itself a smart contract, we can programmatically facilitate every lock being eventually - and even quickly - resolved.  Recalling the motivating example of the introduction, a natural case of this would be restricting lock-swaps to be used by a dedicated {\em swap router}, that is, a contract specifically designed to find and execute appropriate sequences of swaps.
\end{itemize}

\section{Acknowledgements}
{We wish to thank Elie Naba, Burak Can, and Omri Ross for valuable comments on earlier version of this paper. Support by the Center for Blockchains and Electronic Markets, funded by the Carlsberg Foundation under grant no. CF18-1112, is gratefully acknowledged. Pourpouneh also gratefully acknowledges financial support by the Carlsberg Foundation Visiting Fellowships under grant no. CF21-0515.

\bibliography{main}
\bibliographystyle{plain}
\newpage
\section*{Appendix A}
\begin{python}
#
# Python3 prototype of contract for retroactive liquidity pool
#
  
import math
from random import randint
from random import seed

def soft_geq(a, b):
    return a * 1.000001 >= b

def raise_power(a, b):
    if b == 0:
        return 1
    else:
        return a * raise_power(a, b - 1)

def is_even(a):
    return 2 * (a // 2) == a

class Liquiditytokens:
    def __init__(self, no_tokens, product_promise):
        self.no_tokens = no_tokens
        self.product_promise = product_promise

    def __str__(self):
        return (
            "Liquiditytoken portion of "
            + str(self.no_tokens)
            + " tokens,  guaranteeing product "
            + str(self.product_promise)
        )

class Treasury:
    def __init__(self):
        self.treasury = {}

    def add(self, asset, delta):
        if asset in self.treasury:
            self.treasury[asset] += delta
        else:
            self.treasury[asset] = delta

        # Asserting asset positivity:
        assert self.treasury[asset] > 0

    def holding(self, asset):
        if asset in self.treasury:
            return self.treasury[asset]
        else:
            return 0

class SwapLock:
    def __init__(self, delta_a, delta_b):
        self.delta_a = delta_a
        self.delta_b = delta_b

    def __str__(self):
        return "Delta a = " + str(self.delta_a) + ", Delta b =" + str(self.delta_b)

class TraceEntry:
    def __init__(self):
        return

class UnresolvedLock(TraceEntry):
    def __init__(self, lock: SwapLock):
        super().__init__()
        self.lock = lock

    def __str__(self):
        return "Unresolved lock: " + str(self.lock)

class UnresolvedProvide(TraceEntry):
    def __init__(self, delta_a, delta_b):
        super().__init__()
        self.delta_a = delta_a
        self.delta_b = delta_b

    def __str__(self):
        return (
            "Unresolved provide: Delta a = "
            + str(self.delta_a)
            + ", Delta b = "
            + str(self.delta_b)
        )

class UnresolvedReclaim(TraceEntry):
    def __init__(self, tokens: Liquiditytokens):
        super().__init__()
        self.liquiditytokens = tokens

    def __str__(self):
        return "Unresolved reclaim: " + str(self.liquiditytokens)

class ResolvedLocks(TraceEntry):
    def __init__(self, delta_a, delta_b):
        super().__init__()
        self.delta_a = delta_a
        self.delta_b = delta_b

    def __str__(self):
        return (
            "Resolved lock(s): Delta a = "
            + str(self.delta_a)
            + ", Delta b = "
            + str(self.delta_b)
        )

class LiquidityPool:
    def __init__(self, delta_a, delta_b):
        assert delta_a * delta_b > 0
        self.treasury = Treasury()
        self.treasury.add("A", delta_a)
        self.treasury.add("B", delta_b)
        self.a0 = delta_a
        self.b0 = delta_b
        self.events = []
        self.initialTokens = Liquiditytokens(1, delta_a * delta_b)
        self.no_liquiditytokens_minted = 1
        self.liquiditytoken_portions_sent = [self.initialTokens]

    def __str__(self):
        allentries = "#unresolved locks = " + str(self.no_unresolved_locks())
        c = 0
        for e in self.events:
            allentries = allentries + "\n" + str(c) + ": " + str(e)
            c += 1
        if not (allentries == ""):
            allentries = "\n" + allentries + "\n"
        statstring = (
            "\nLiquiditytokens minted and unburnt: "
            + str(self.no_liquiditytokens_minted)
            + "\nLiquiditytoken portions sent to providers: "
        )
        c = 0
        for p in self.liquiditytoken_portions_sent:
            statstring = statstring + "\n" + str(c) + ": " + str(p)
            c += 1
        zerostring = (
            "\nPool state: \n\na0 = "
            + str(self.a0)
            + ", b0 = "
            + str(self.b0)
            + ", Product = "
            + str(self.a0 * self.b0)
        )
        currentstring = (
            "\nActually stored in pool (real assets!) a = "
            + str(self.treasury.holding("A"))
            + ", b = "
            + str(self.treasury.holding("B"))
        )
        return zerostring + allentries + currentstring + statstring + "\n\n"

    def add_event(self, e: TraceEntry):
        self.events = self.events + [e]
        self.maintain_resolved_locks_invariant_at(len(self.events) - 1)
        self.maintain_first_position_invariant()

    def no_unresolved_locks(self):
        c = 0
        for i in range(0, len(self.events)):
            e = self.events[i]
            if isinstance(e, UnresolvedLock):
                c += 1
        return c

    def position_of(self, lock: SwapLock):
        for i in range(0, len(self.events)):
            e = self.events[i]
            if isinstance(e, UnresolvedLock) and e.lock == lock:
                return i
        return -1

    def merge_resolved_locks_at(self, position):
        assert isinstance(self.events[position], ResolvedLocks)
        while len(self.events) >= position + 2 and isinstance(
            self.events[position + 1], ResolvedLocks
        ):
            self.events[position].delta_a += self.events[position + 1].delta_a
            self.events[position].delta_b += self.events[position + 1].delta_b
            del self.events[position + 1]

    def maintain_resolved_locks_invariant_at(self, position):
        if not (isinstance(self.events[position], ResolvedLocks)):
            return
        elif position == 0:
            self.merge_resolved_locks_at(0)
        elif isinstance(self.events[position - 1], ResolvedLocks):
            self.maintain_resolved_locks_invariant_at(position - 1)
        else:
            self.merge_resolved_locks_at(position)

    def handle_resolved_locks_at_first_position(self):
        resolved = self.events.pop(0)
        self.a0 = self.a0 + resolved.delta_a
        self.b0 = self.b0 + resolved.delta_b
        assert self.a0 > 0
        assert self.b0 > 0

    def handle_unresolved_provide_at_first_position(self):
        provide = self.events.pop(0)
        new_a0 = self.a0 + provide.delta_a
        new_b0 = self.b0 + provide.delta_b
        delta = new_a0 * new_b0 / (self.a0 * self.b0) - 1
        epsilon = math.sqrt(1 + delta) - 1
        promise = provide.delta_a * provide.delta_b
        no_tokens = epsilon * self.no_liquiditytokens_minted
        liquiditytokens = Liquiditytokens(no_tokens, promise)
        self.no_liquiditytokens_minted += no_tokens
        self.liquiditytoken_portions_sent =
            self.liquiditytoken_portions_sent + [liquiditytokens]
        self.a0 = new_a0
        self.b0 = new_b0

    def handle_unresolved_reclaim_at_first_position(self):
        reclaim = self.events.pop(0)
        no_tokens = reclaim.liquiditytokens.no_tokens
        pool_fraction = no_tokens / self.no_liquiditytokens_minted
        withdrawn_a = self.a0 * pool_fraction
        withdrawn_b = self.b0 * pool_fraction
        self.no_liquiditytokens_minted -= no_tokens

        # Asserting product preservation:
        assert soft_geq(withdrawn_a * withdrawn_b, reclaim.liquiditytokens.product_promise)

        self.a0 -= withdrawn_a
        self.b0 -= withdrawn_b
        self.treasury.add("A", -withdrawn_a)
        self.treasury.add("B", -withdrawn_b)

    def maintain_first_position_invariant(self):
        if self.events == []:
            return
        if isinstance(self.events[0], UnresolvedLock):
            return
        if isinstance(self.events[0], ResolvedLocks):
            self.handle_resolved_locks_at_first_position()
        elif isinstance(self.events[0], UnresolvedProvide):
            self.handle_unresolved_provide_at_first_position()
        else:
            assert isinstance(self.events[0], UnresolvedReclaim)
            self.handle_unresolved_reclaim_at_first_position()
        self.maintain_first_position_invariant()

    def cancel_lock_at(self, position):
        assert isinstance(self.events[position], UnresolvedLock)
        del self.events[position]
        if position < len(self.events) and isinstance(
            self.events[position], ResolvedLocks
        ):
            self.maintain_resolved_locks_invariant_at(position)
        self.maintain_first_position_invariant()

    def execute_lock_at(self, position):
        assert isinstance(self.events[position], UnresolvedLock)
        ulock = self.events[position].lock
        rlock = ResolvedLocks(ulock.delta_a, ulock.delta_b)
        self.treasury.add("A", ulock.delta_a)
        self.treasury.add("B", ulock.delta_b)
        self.events[position] = rlock
        self.maintain_resolved_locks_invariant_at(position)
        self.maintain_first_position_invariant()

    def all_possible_allocations(self):
        k = self.no_unresolved_locks()
        cases = raise_power(2, k)
        for rho in range(0, cases):
            my_rho = rho
            a = self.a0
            b = self.b0
            no_tokens = self.no_liquiditytokens_minted
            for e in self.events:
                if isinstance(e, UnresolvedLock):
                    if is_even(my_rho):
                        a += e.lock.delta_a
                        b += e.lock.delta_b
                    my_rho = my_rho // 2
                elif isinstance(e, ResolvedLocks):
                    a += e.delta_a
                    b += e.delta_b
                elif isinstance(e, UnresolvedProvide):
                    new_a = a + e.delta_a
                    new_b = b + e.delta_b
                    delta = new_a * new_b / (a * b) - 1
                    epsilon = math.sqrt(1 + delta) - 1
                    no_tokens += epsilon * no_tokens
                    a = new_a
                    b = new_b
                else:
                    assert isinstance(e, UnresolvedReclaim)
                    burn_tokens = e.liquiditytokens.no_tokens
                    pool_fraction = burn_tokens / no_tokens
                    no_tokens -= burn_tokens
                    a *= 1 - pool_fraction
                    b *= 1 - pool_fraction
            yield a, b

    # Below are the standard external operations

    def lock_swap_a_to_b(self, delta_a):
        smallest_delta_b = self.treasury.holding("B")
        for a, b in self.all_possible_allocations():
            delta_b = b - a * b / (a + delta_a)
            if delta_b < smallest_delta_b:
                smallest_delta_b = delta_b
        lock = SwapLock(delta_a, -smallest_delta_b)
        self.add_event(UnresolvedLock(lock))
        return lock

    def lock_swap_b_to_a(self, delta_b):
        smallest_delta_a = self.treasury.holding("A")
        for a, b in self.all_possible_allocations():
            delta_a = a - a * b / (b + delta_b)
            if delta_a < smallest_delta_a:
                smallest_delta_a = delta_a
        lock = SwapLock(-smallest_delta_a, delta_b)
        self.add_event(UnresolvedLock(lock))
        return lock

    def cancel(self, lock):
        position = self.position_of(lock)
        self.cancel_lock_at(position)

    def execute(self, lock):
        position = self.position_of(lock)
        self.execute_lock_at(position)

    def provide(self, delta_a, delta_b):
        self.treasury.add("A", delta_a)
        self.treasury.add("B", delta_b)
        self.add_event(UnresolvedProvide(delta_a, delta_b))

    def reclaim(self, t: Liquiditytokens):
        self.add_event(UnresolvedReclaim(t))

#Below is code testing the pool with random operations 

seed(252352)
epoch = 1
epoch_length = 10000
while True:
    print("Epoch #", epoch, " of ", epoch_length, " operations:")
    p = LiquidityPool(100, 100)
    locks = []
    do_print = True
    epoch = epoch + 1

    for j in range(0, epoch_length):
        print("Operation #", j, " of epoch", epoch - 1)
        if do_print:
            print(p)
        c = randint(1, 8)
        do_print = False
        if c == 1 and (p.no_unresolved_locks() < 18):
            print("lockswapping A to B")
            locks = locks + [p.lock_swap_a_to_b(randint(1, 150))]
            do_print = True
        if c == 2 and (p.no_unresolved_locks() < 18):
            print("lockswapping B to A")
            locks = locks + [p.lock_swap_b_to_a(randint(1, 150))]
            do_print = True
        if c == 3:
            if not (locks == []):
                print("Cancelling a lock")
                d = randint(0, len(locks) - 1)
                lock = locks.pop(d)
                p.cancel(lock)
                do_print = True
        if c == 4 or c == 7:
            if not (locks == []):
                print("Executing a lock")
                d = randint(0, len(locks) - 1)
                lock = locks.pop(d)
                p.execute(lock)
                do_print = True
        if c == 5:
            print("Providing liquidity")
            p.provide(randint(1, 150), randint(1, 150))
            do_print = True
        if c == 6 or c == 8:
            if len(p.liquiditytoken_portions_sent) >= 2:
                print("Reclaiming liquidity")
                d = randint(1, len(p.liquiditytoken_portions_sent) - 1)
                t = p.liquiditytoken_portions_sent.pop(d)
                p.reclaim(t)
                do_print = True
\end{python}
\end{document}